\documentclass[a4paper,12pt,tightenlines,notitlepage,nofootinbib]{revtex4-2}
\usepackage{physics}
\usepackage{amsmath,amssymb}
\usepackage{graphicx}
\usepackage{newtxtext,newtxmath}
\usepackage[normalem]{ulem}
\usepackage[x11names]{xcolor}
\usepackage[unicode,colorlinks,citecolor=Blue3,linkcolor=Blue3,bookmarks=true]{hyperref}

\usepackage{empheq}
\newenvironment{eqw}{\begin{equation} \begin{aligned}}
    {\end{aligned}    \end{equation}}
\newenvironment{eqw*}{\begin{equation*} \begin{aligned}}
    {\end{aligned}    \end{equation*}}

\DeclareMathOperator*{\sign}{sign}
\DeclareMathOperator*{\Real}{Re}
\DeclareMathOperator*{\Imag}{Im}
\DeclareMathOperator*{\arcsh}{arcsinh}

\DeclareMathOperator*{\Arcsinh}{arcsh}
\newcommand{\mean}[1]{\langle{#1}\rangle}

\newcommand{\cint}{\int\limits_{\mathbb{C}}}
\newcommand{\tbeta}{\tilde\beta}
\newcommand{\jes}{{j_1 j_2}}

\begin{document}
\title{Effective algorithms for calculation of quasiprobability distributions of bright ``banana'' states}

\author{Boulat Nougmanov}


\affiliation{Russian Quantum Center, Skolkovo IC, Bolshoy Bulvar 30, bld. 1, Moscow, 121205, Russia}

\affiliation{Moscow Institute of Physics and Technology, 9 Institutsky Lane, Dolgoprudny, 141701, Russia}

\begin{abstract}
Non-Gaussian quantum states, described by negative valued Wigner functions, are important both for fundamental tests of quantum physics and for emerging quantum information technologies. One of the promising ways of generation of the non-Gaussian states is the use of the cubic (Kerr) optical non-linearity, which produces the characteristic banana-like shape of the resulting quantum states. However, the Kerr effect in highly transparent optical materials is weak. Therefore, big number of the photons in the optical mode ($n\gtrsim10^6$) is necessary to generate an observable non-Gaussianity. In this case, the direct approach to calculation of the Wigner function becomes extremely computationally expensive.

In this work, we develop quick algorithms for computing the Husimi and Wigner quasiprobability  functions of these non-Gaussin states by means of the Kerr nonlinearity. This algorithm can be used for any realistic values of the photons number and the non-linearity.
\end{abstract}

\maketitle

\section{Introduction}

One of the priority tasks of quantum technologies is to design and build a universal quantum computer, allowing to provide an exponential speedup over classical computing. The promising approach to implementation of the quantum computing is based on the use of entangled modes of light (qumodes) and the corresponding continuous variables (quadrature amplitudes) of these modes \cite{Lloyd_PRL_82_1784_1999, Braunstein_RMP_77_513_2005, Weedbrook_RMP_84_621_2012}. The advantage of this approach is its very high degree of scalability \cite{Pfister_JoPB_53_012001_2019}

The toolbox for the continuous variables quantum computing includes linear optical elements, such as beamsplitters and $\chi^{(2)}$-nonlinearity based optical parametric amplifiers, described by the Hamiltonians which are quadratic in the quadrature amplitudes. An important feature of these elements is that they transform Gaussian quantum states (that is the ones described by the Guassian quasi-probability distributions, see Sec.\,\ref{sec:WH}) again to Gaussian ones. At the same time, it is known that any setup which uses only Gaussian quantum states can be effectively simulated by the classical computer \cite{Bartlett_88_097904_2002, Mari_PRL_109_230503_2012}. Therefore, the key condition for achieving quantum supremacy is the use on non-linear elements and, correspondingly, non-Gaussian quantum states.

In the optical band, the most promising type of nonlinearity is the qubic ($\chi^{(3)}$), or Kerr one (the higher order optical nonlinearities are negligibly small, see {\it e.g.} \cite{ekvall2001studies}). In the rotating wave picture, the Hamiltonian of optical mode with in such a nonlinearity is equal to \cite{book93397576}: 
\begin{eqw}\label{H_Kerr}
  \hat{H} = -\hbar\gamma\hat{n}\left(\hat{n}-1\right),
\end{eqw}
where $\gamma$ is the nonlinearity parameter proportional to $\chi^{(3)}$ and inversely proportional to the mode volume \cite{Kitagawa_PRA_34_3974_1986, 20a1BaKhStMaBi}, and $\hat{n}$ is the operator of the photons number in the mode. Since this Hamiltonian commutes with $\hat{n}$, the photon number distribution does not change during evolution. At the same time, the optical phase is modified proportionally $\hat{n}$ (the self phase modulation effect, SPM). As a result, the quantum state of the optical mode acquires the characteristic non-Gaussian banana-like shape \cite{Kitagawa_PRA_34_3974_1986}. The degree on the non-Gaussianity is defined by the dimensionless parameter
\begin{equation}
  \Gamma = \gamma\tau \,.
\end{equation}
where $\tau$ is the evolution time that should be much shorter than the intrinsic relaxation time of the mode $\tau^*$. The values of
\begin{equation}
  \Gamma \gtrsim \frac{1}{\mean{n}^{5/6}} \,,
\end{equation}
where $\mean{n}$ is the mean photon number, correspond to the photon number uncertainties
\begin{equation}
  \Delta n \lesssim \mean{n}^{1/3}
\end{equation}
that are unreachable by linear operations (squeezing). Therefore they can be considered as the ``witness'' of the non-Gaussianity \cite{Bondurant_PRD_30_2548_1984, Kitagawa_PRA_34_3974_1986, 20a1BaKhStMaBi}.

Unfortunatetely, the Kerr effect in highly transparent optical materials is weak. This problem can be alleviated by concentration of the optical power in the small volume of the mode. The promising way to overcome this problem is the usage of optical ring resonators \cite{89a1BrGoIl, Strekalov_JOptics_18_123002_2016}, which combine the small volume of the mode with very high quality (more than $10^{11}$ in crystalline microresonators \cite{Savchenkov_OE_15_6768_2007} and more than $10^9$ in on-chip ones \cite{Kippenberg_PRL_93_083904_2004, Wu_OL_45_5129_2020}). However, even in the best modern microresonators,
\begin{equation}
  \Gamma \lesssim 10^{-6} \,,
\end{equation}
see {\it e.g.} esimates in Ref.\,\cite{20a1BaKhStMaBi}, which translates to
the values
\begin{equation}
  \mean{n} \gtrsim 10^6 \,.
\end{equation}

For the theoretical study of the ``banana'' states, it is necessary to calculate their quasi-probability distributions, namely the  Husimi and Wigner functions (see Sec.\,\ref{sec:WH}). However, here the computational problem arises. While the Husimi function can be calculated directly using its definition without significant difficulties \cite{rosiek2022enhancing}, calculation of the Wigner function is much more involved.

Two methods of numerical calculation of the Wigner function were typically used in literature: by solving the differential equation on its evolution over time, or directly by calculating it for each point. As the results of Ref.\,\cite{stobinska2008wigner} show, the known formulas for direct calculation turn out to be inapplicable for large $\mean{n}$. In addition to causing overflows quickly, they also require about $\mean{n}^2$ summands, which makes them very computationally expensive.

In this paper, we build an alternative approach to the computation of the Husimi and Wigner functions based on asymptotic estimates that utilize the smallness of $\Gamma$ and the large value of $\mean{n}$. The paper is organized as follows. In Sec.\,\ref{sec:WH}, we briefly overview the main features of the Husimi and Wigner functions. Then, in Sections \ref{sec:Husimi} and \ref{sec:Wigner}, we develop the fast algorithms for  numerical computing of the Husimi and Wigner function, respectively. In Sec.\,\ref{sec:conclusion}, we summarize the main results of this paper.

\section{Introduction to Husimi and Wigner functions}\label{sec:WH}

The quantum quasi-probability distributions \cite{Wigner_PR_40_749_1932, Husimi_PPMSJ3_22-264_1940, Glauber_PR_131_2766_1963, Cahill_PR_177_1882_1969} serve as conveninent substitutes for the joint probability distributions for the positions and momenta of quantum objects. They have one-to-one correspondence with the respective quantum states and can be experimentally reconstructed using the quantum tomography procedure \cite{Vogel_PRA_40_2847_1989}.

It is convenient to introduce the quantum quasi-probability distributions, using the s-parametrized characteristic function, defined as follows \cite{Cahill_PR_177_1882_1969}:
\begin{equation}\label{C_s}
  C(z,s) = \Tr\bigl[\hat{\rho}e^{i(z^*\hat{a} + z\hat{a}^\dag)}\bigr]e^{s|z|^2/2} \,,
\end{equation}
where $\hat{a}$, $\hat{a}^\dag$ are the annihilation and creation operator, $\hat{\rho}$ is the density operator, $z$ is a complex number, and $s$ is a real parameter with $|s|\le1$. The Wigner function \cite{Wigner_PR_40_749_1932} is equal to the inverse Fourier transform of the symmetrical ordered ($s=0$) characteristic function
\begin{equation}\label{Wigner}
  W(\beta) = \int C(z,0)e^{-i(z^*\beta + z\beta^*)}\,\frac{d^2z}{\pi^2} \,,
\end{equation}
where
\begin{equation}
  \beta = \frac{x + ip}{\sqrt{2}}
\end{equation}
and $x$, $p$ are the normalized position and momentum. The distinctive feature of the Wigner functaion is that in the Gaussian case, it is exactly equal to the corresponding classical probability distribution. However, the Wigner functions of all other (non-Gaussian) pure quantum states take negative values \cite{Hudson_RMP_6_249_1974}. It is this feature actually makes the these states ``truly quantum''. The Wigner function can be measured indirectly using homodyne detection, which makes the experimental verification of these results relatively simple \cite{Schleich2001}.

The inverse Fourier transform of the anti-normal ordered ($s=-1$) characteristic function gives the Husimi $Q$-function \cite{Husimi_PPMSJ3_22-264_1940}:
\begin{equation}\label{Husimi}
  Q(\beta) = \int C(z,-1)e^{-i(z^*\beta + z\beta^*)}\,\frac{d^2z}{\pi^2}
    = \frac{1}{\pi}\bra{\beta}\hat{\rho}\ket{\beta}\,,
\end{equation}
where $\ket{\beta}$ is a coherent state. It is non-negative everywhere and can be obtained from the Wigner function by the Gaussian blurring operation.

Yet another quasi-probability distribution, the Glauber $P$-function \cite{Glauber_PR_131_2766_1963}, correspond to the normal ordering case of $s=1$. However, it is highly singular for all pure quantum states except of the ground and coherent ones. Therefore, we will not consider it here.

\section{Calculation of Husimi function}\label{sec:Husimi}

Evolution of the initial coherent quantum state $\ket{\alpha}$ with the Hamiltonian \eqref{H_Kerr} during the time $\tau$ gives the following wave function:
\begin{equation}
  \ket{\psi} = \sum_{n=0}^\infty\frac{\alpha^n e^{i\Gamma n(n-1)}}{\sqrt{n!}}\ket{n} .
\end{equation}
Using then the definition \eqref{Husimi}, we obtain the following equation for the corresponding Husimi function:
\begin{equation}\label{Husimi by def}
    Q(\beta) = \frac{\abs{\braket{\beta}{\psi}}^2}{\pi}
    = \frac{e^{-\abs{\alpha}^2-\abs{\beta}^2}}{\pi}\left|
	\sum\limits_{n = 0}^{\infty}\dfrac{\left(\alpha \beta^*\right)^n e^{i \Gamma n(n-1)}}{n!}\right|^2
\end{equation}

Note that $\alpha$ and $\beta$ appear in the Husimi function series only in combination
$A=\alpha\beta^*e^{-i\Gamma}$. Therefore, let us introduce the function
\begin{equation}\label{F_def}
  F(A) = \sum\limits_{n=0}^{\infty} \frac{A^n}{n!}e^{i\Gamma n^2} \,.
\end{equation}
Using this notation, the Husimi function \eqref{Husimi by def} can be expressed as follows::
\begin{equation}\label{QtoF}
  Q(\beta) = \frac{e^{-\abs{\alpha}^2-\abs{\beta}^2}}{\pi}\abs{F\left(\alpha\beta^*e^{-i\Gamma}\right)}^2
\end{equation}
The nonlinear evolution factor in Eq.\,\eqref{F_def} can be presented as follows:
\begin{equation}
    e^{i\Gamma n^2} = \frac{e^{i\frac{\pi}{4}\sign \Gamma}}{2\sqrt{\pi|\Gamma|}}
      \fint\limits_{-\infty}^{\infty} \exp\left(-\frac{i z^2}{4\Gamma} + i n z\right)dz \,,
\end{equation}
where $\fint$ stands for the Cauchy principal value integral. Therefore,
\begin{equation}\label{F_of_A}
  F(A) = \frac{e^{i\frac{\pi}{4}\sign \Gamma}}{2\sqrt{\pi|\Gamma|}}
    \fint\limits_{-\infty}^{\infty} \exp\left(\frac{f(z)}{2\Gamma}\right) dz \,,
\end{equation}
where
\begin{subequations}
  \begin{gather}
    f(z) = \frac{z^2}{2i} + i Z e^{iz} \,,\\
    Z = -2i A \Gamma \,.
  \end{gather}
\end{subequations}

\begin{figure}
    \centering
    \includegraphics{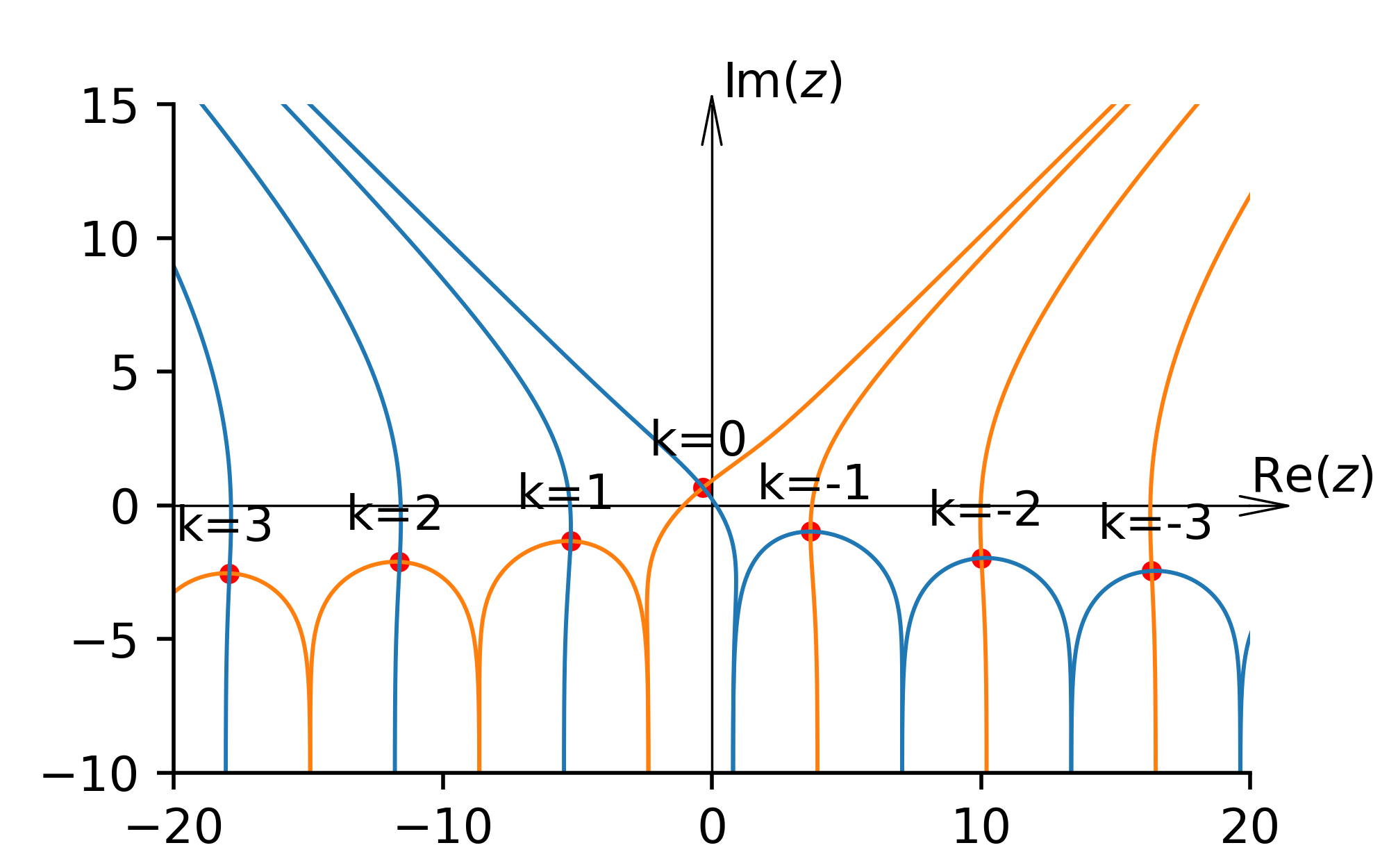}
    \caption{Constant phase lines for the integral \eqref{F_of_A} at $Z=1+i$. Blue indicates the lines of the steepest descent $\gamma_k$ at $\Gamma < 0$, orange --- at $\Gamma > 0$. The red points of intersection of the lines are the points of the cusp $f'(z_k)=0$.}\label{saddle_points}
\end{figure}

The integral in Eq.\,\eqref{F_of_A} can be calculated using the saddle point method \cite{fedoryuk1977saddle}. In our case, the number of pass points $z_k$ is infinite ($k\in\mathbb{Z}$). They can be numbered according to the numbers of branches of the Lambert $W$-function:
\begin{eqw}\label{z_k def}
     f'(z_k) =0 \Rightarrow i Z e^{i z_k} = z_k \Rightarrow   z_k = i W_k(Z)
\end{eqw}
The schematic positions of the saddle points $z_k$ and the constant phase lines $\gamma_k$ passing through them is shown in Fig.\,\ref{saddle_points}. Note that there are two constant phase lines for each of $z_k$: one of them corresponds to the line of steepest descent at $\Gamma > 0$, and the other --- at $\Gamma < 0$.

A detailed analysis at how to deform the integration contour and account for residual terms using the Campbell–Froman–Walles–Wojdylo (CFWW) formula \cite{bleistein1975asymptotic} is done in Appendix \ref{appA}. It is shown there that the deformed contour passes through only half of all $z_k$, which gives to the following result:
\begin{eqw}\label{sum_k}
    \fint\limits_{-\infty}^{\infty}\exp\left(\frac{f(z)}{2\Gamma}\right)dz = \sum\limits_{k=0}^{-\sign\Gamma \cdot \infty} \sqrt{\frac{4\Gamma}{{-i-z_k}}}\exp\left(\frac{ f(z_k)}{2\Gamma} \right)\left(1+O(\Gamma)\right)
\end{eqw}

\begin{figure}
    \centering
    \includegraphics[width=0.75\textwidth]{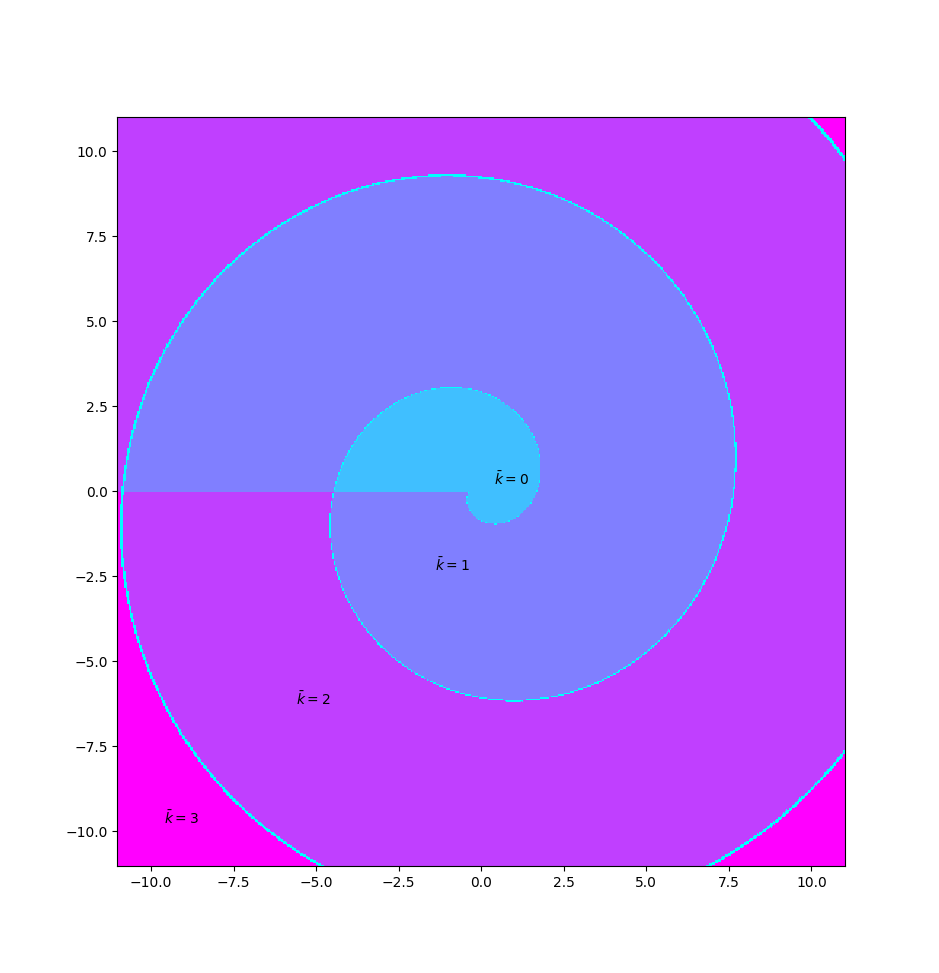}
    \caption{
    Every point $Z$ in the complex plane has an associated $\bar{k}$ that makes the primary contribution to the sum \eqref{sum_k}. The narrow blue line denotes small set of points where $\abs{\Real f(z_k) - \Real f(z_{k+1})}<2\abs{\Gamma}$. In this plot, the value of $\Gamma=0.01$ is used. With the decreases of $\Gamma$, the measure of points $Z$ at which two summands ($\bar{k}$ and $\bar{k}+1$) have to be considered in the sum \eqref{sum_k} decreases as well.}\label{where_best_k_bar2}
\end{figure}

\begin{figure}
    \centering
    \includegraphics[width=0.6\textwidth]{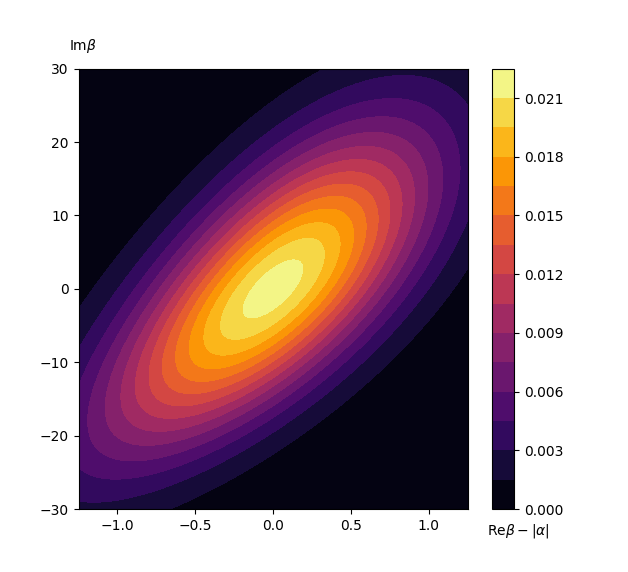}
    \caption{Contour plot of the Husimi function at $\abs{\alpha}=2700$, $\Gamma=10^{-6}$. The choice of $\arg{\alpha}$ was determined from the condition that the peak value of the quasidistribution is on the real axis. The slightly asymmetric twist is clearly visible in the picture.}
    \label{figure_Husimi}
\end{figure}

This equation is still not convenient for numerical calculations, because it contains an infinite sum. However, due to the smallness of $\Gamma$, the term $\abs{\exp\left(\frac{ f(z_k)}{2\Gamma} \right)}$, as a function of $k$, has a very narrow Gaussian profile. It is shown in Appendix \ref{appB}, that due to this reason, that out of the whole sum in the equation \eqref{sum_k}, only one or two summands should be kept, as shown, in Figure \ref{where_best_k_bar2}, and in most cases, it is sufficient to keep only one summand with the number
\begin{equation}
  \bar k = -\sign \Gamma \left[ \frac{1}{2\pi}\left(2\left|\Gamma \alpha\beta^*\right| + \left|\arg{\left(\Gamma \alpha\beta^* e^{-i\Gamma}\right)} + \frac{\pi}{2}\sign \Gamma\right|\right)\right],
\end{equation}
where $[...]$ means rounding to the nearest integer. As a result, we obtain the following closed equation for the Husimi function:
\begin{equation}\label{Qgood}
  Q(\beta) \approx \frac{
      \exp\left\{
          -\frac{1}{2\Gamma}\Imag\left[
              1+W_{\bar k}\left(-2i\alpha\beta^* \Gamma e^{-i\Gamma}\right)
            \right]^2
          -\abs{\alpha}^2 - \abs{\beta}^2
        \right\}
    }{\pi^2\abs{1+W_{\bar k}\left(-2i\alpha\beta^* \Gamma e^{-i\Gamma}\right)}} \,.
\end{equation}
A characteristic example of the Husimi function, calculated using this equation, is shown in Fig.\,\ref{figure_Husimi}.

\section{Wigner function}\label{sec:Wigner}

\begin{figure}
    \centering
    \includegraphics[width=0.75\textwidth]{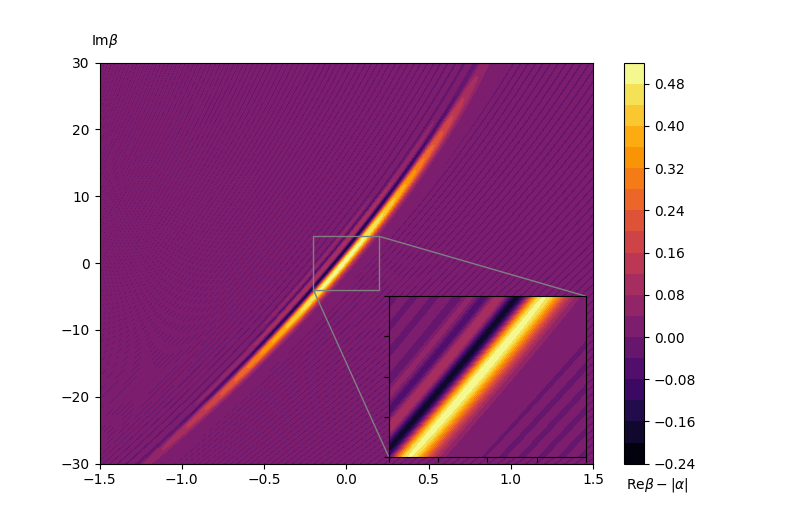}
    \caption{Contour plot of the Wigner function under the same parameters as in figure \ref{figure_Husimi}. As can be seen, the meaningful part of the Wigner function is stretched over a wider sector,  according to \cite{Kitagawa_PRA_34_3974_1986}. Long narrow valleys of negativity are clearly visible.}
    \label{figure_Wigner}
\end{figure}
The basic idea for finding the values of the Wigner function is to relate it to the Husimi function using Eqs.\,\eqref{C_s}, \eqref{Wigner}, and \eqref{Husimi}:.
\begin{multline}\label{FourierC}
   W(\beta) = \int\limits_{\mathbb{C}} C(z,-1)e^{|z|^2/2-i(z^*\beta + z\beta^*)}\frac{d^2z}{\pi^2} \\
    = \cint\frac{d^2z}{\pi^2} e^{-i(z^*\beta + z \beta^*)} e^{|z|^2/2}
		{\cint d^2\tbeta\,e^{i(z^*\tbeta + z \tbeta^*)} Q(\tbeta)} \,.
\end{multline}
However, for the numerical calculation, we cannot use directly in this equation the already obtained approximation for the Husimi function \eqref{Qgood} due to the exponential factor $e^{\abs{z}^2/2}$.
At the same time, if $\Gamma/(2\pi)$ is a rational number:
\begin{equation}
  \frac{\Gamma}{2\pi} = \frac{k}{n} \,,
\end{equation}
where $k$, $n$ are integers, then the following closed expression with a finite number of summands exists for the function $F$:
\begin{eqw}\label{Fexp}
    F(A) = \frac{\sum\limits_{j=1}^n\exp\left(-i\Gamma j^2 + A e^{2ij\Gamma}\right)}{\sum\limits_{j=1}^n\exp\left(-i\Gamma j^2\right)} \,,
\end{eqw}
see Appendix \ref{appC}. Any particular gamma observed in an experiment will almost surely be irrational. However, since the rational numbers are dense in the set of real numbers, this formula can approximate the actual $\gamma$ dependence as accurately as necessary with a proper choice of $k$ and $n$.

 The representation \eqref{Fexp} together with \eqref{QtoF} is very convenient, since it reduces calculation the Fourier transforms  \eqref{FourierC} to computation of Gaussian integrals,  see  Appendix \ref{appD}. The result is the following formula, which is valid for arbitrary $\Gamma\in\mathbb{R}$:
\begin{eqw}\label{Q_to_W_for_banan}
    W(\beta) = 2e^{2|\beta|^2}\sum\limits_{m=0}^{\infty} \frac{\left(-|\alpha|^2\right)^m}{m!}Q(2\beta e^{-2im\Gamma })
\end{eqw}

This form is still can not be used directly in numerical calculations. In the most meaningful case $\abs{\beta}\sim\abs{\alpha}$ it may be shown that as well as for the coherent state $Q(2\beta e^{-2im\Gamma })\sim e^{-\abs{\alpha}^2}$ for some $m$. The maximum of $\frac{|\alpha|^{2m}}{m!}$ have the  order of magnitude of $e^{\abs{\alpha}^2}$. Therefore, to get values $W(\beta)\sim 1$ we need to sum numbers of order $1$ with the accuracy of order $\sim e^{-2\abs{\alpha}^2}$, which is unattainable at large $\abs{\alpha}$.

In order to evade this problem, let us decompose the expression \eqref{Q_to_W_for_banan} into a Fourier series in $\Phi_0 = \arg\left(\alpha\beta^* e^{-i\Gamma}\right)$:
\begin{eqw}\label{Fourier W}
     W(\beta) = \frac{2}{\pi}e^{-2\left(\abs{\beta} - \abs{\alpha}\right)^2}
     \sum\limits_{k=-\infty}^{+\infty} e^{ik\Phi_0} I_{k}\left(4\abs{\alpha\beta^*}e^{ik\Gamma}\right)
     \exp\left(\abs{\alpha}^2 (1-e^{2ik\Gamma})-4\abs{\alpha\beta^*}\right) .
\end{eqw}
Here we can use the asymptotics for the modified Bessel function $I_k$, which are valid both for small and large values $k$. Using Eq.\,\eqref{I_bateman_asymptotic} and assuming that $k\Gamma\in\left(-\frac{\pi}{2}, \frac{\pi}{2}\right)$, we obtain the following equation for the Wigner function:
\begin{multline}\label{W_final}
    W(\beta) \approx \frac{1}{\pi}\sqrt{\frac{2}{\pi}}\sum\limits_{k=-k_{\max}}^{+k_{\max}}
    \left(\left(4\abs{\alpha\beta^*}e^{ik\Gamma}\right)^2+k^2\right)^{-\frac{1}{4}}
    \exp\Bigg{(}-2\left(\abs{\beta} - \abs{\alpha}\right)^2 + ik\arg\left(\alpha\beta^* e^{-i\Gamma}\right) \\
    + \abs{\alpha}^2\left(1-e^{2ik\Gamma}\right) - 4\abs{\alpha\beta^*}
    + \sqrt{\left(4\abs{\alpha\beta^*}e^{ik\Gamma}\right)^2+k^2}
    - k\arcsh\left(\frac{k}{4\abs{\alpha\beta^*}e^{ik\Gamma}}\right)\Bigg{)} ,
\end{multline}
where $k_{\max}$ is determined numerically based on the required accuracy.

More details about the Fourier series expansion \eqref{Fourier W}, linear asymptotics of $k_{\max}$ and residual terms of the expansion \eqref{W_final} can be found in the Appendix \ref{appE}.

An example of the Wigner function, calculated using Eq.\,\eqref{W_final}, is shown in Fig.\,\ref{figure_Wigner}.

\section{Conclusion}\label{sec:conclusion}

The fast algorithms developed in this work allow to radically accelerate calculation of quasi-probability distribution of the the mutli-photon ``banana'' states.

The standard algorithm for computing the Husimi function, based on the direct computation of the sum of Poisson distributed terms \eqref{Husimi by def}, require the computing time that scales as $O(\mean{n}^{1/2})$. The algorithm which we developed here, allows one to calculate the Husimi function for $O(1)$ operations.

In the case of the Wigner function, the difference is much more significant.
The direct algorithm requires $O(\mean{n}^2)$ operations and leads to overflow when $\mean{n}$ is large. Our algorithm requires the computing time that scales as $O(\mean{n}^{1/2})$ only.

The software implementation of these algorithms is available on the github \cite{mygit}.

\acknowledgments

I appreciate assistance and guidance from Farit Khalili.
This work was supported by the Russian Science Foundation (project 20-12-00344).

\appendix

\section{Series for Husimi functions}\label{appA}
\subsection{Contour deformation}
The purpose of this section is to show how the integration contour on the complex plane can be deformed using the following function:
\begin{eqw}\label{main_integral}
    &\fint\limits_{-\infty}^{\infty}\exp\left(\frac{f(z)}{2\Gamma}\right) dz = 2\sqrt{\pi|\Gamma|}e^{-i\frac{\pi}{4}\sign \Gamma}\sum\limits_{n=0}^{\infty} \frac{A^n e^{i\Gamma n^2}}{n!}\text{, where }\\
    &f(z) =  \frac{z^2}{2i} + i Z e^{iz}\\
    &Z = -2i A \Gamma = R e^{i\Phi}
\end{eqw}
We will consistently apply the steepest descent method.
\subsubsection{Pass points}
First we find the pass points satisfying the equation $f'(\tilde z)=0$:
\begin{eqw}
    e^{i \tilde z} = \frac{\tilde z}{iZ} \Rightarrow   \tilde z = i W(Z),
\end{eqw}
where $W$ is the $W$-function of the Lambert function. Note that it has an infinite number of branches numbered by integer $k$. Let's index the pass points according to this numbering:
\begin{eqw}
    z_k = i W_k(Z)
\end{eqw}
\begin{figure}[!ht]
    \centering
    \includegraphics[width=0.9\textwidth]{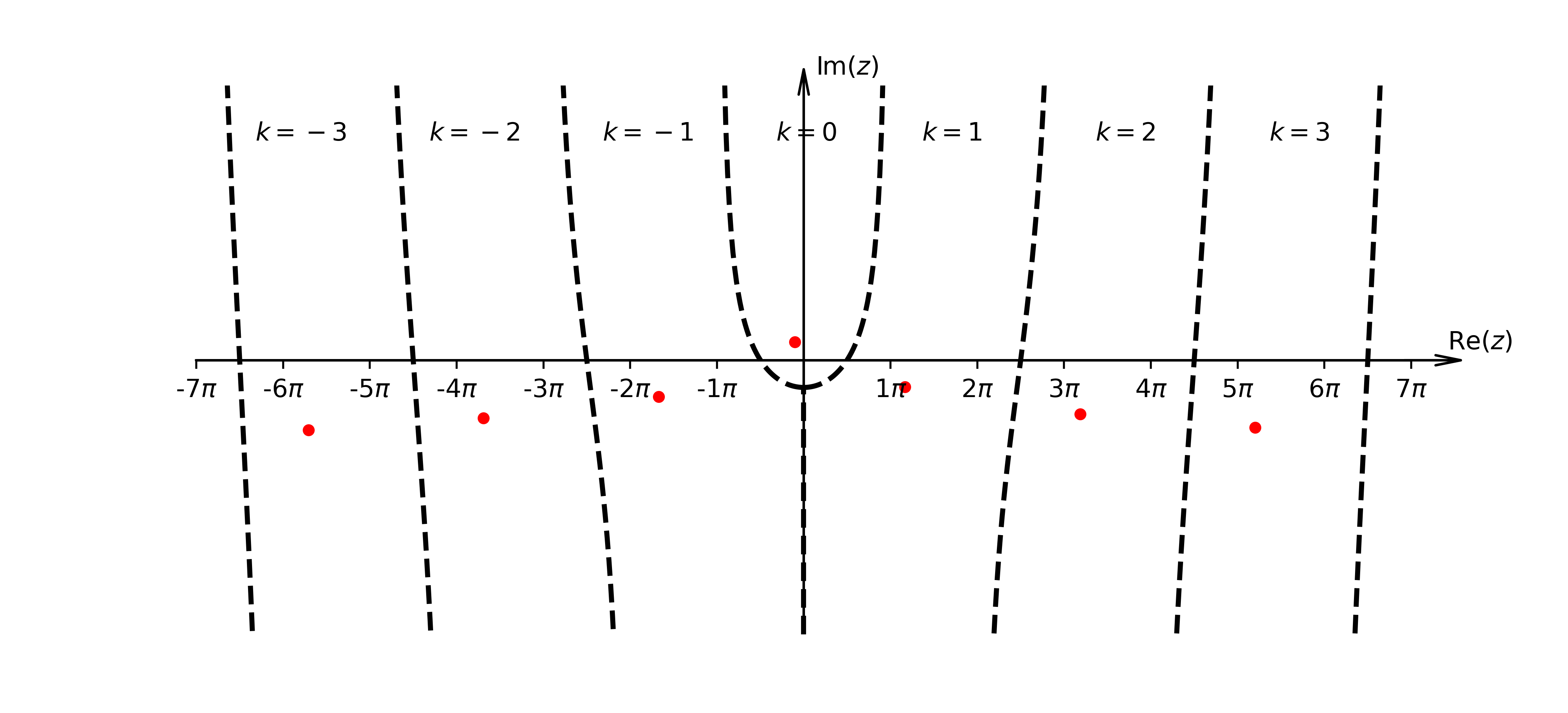}
    \caption{The dotted lines indicate regions where $z_k$ can be located. Red dots mark the positions of $z_k$ at $Z=1+i$.}\label{figure_z_k_position}
\end{figure}
Figure  \ref{figure_z_k_position} illustrates the possible locations of $z_k$. A special case that we will not consider is $Z=-e^{-1}$. By excluding this special $Z$, we can say that all $z_k$ are distinct from each other, are second-order pass points ($f''(z_k)\neq 0$) and every one of them satisfies $z_k \neq -i$.
\subsubsection{Asymptotics of constant phase curves}
Let us consider some point $z_k$. Since $f'(z_k)= 0$ and $f''(z_k)\neq 0$, there are two constant phase curves through $z_k$. One of these is the steepest descent curve and the other is the steepest growth curve. (These curves change when the sign of $\Gamma$ changes.) Let us assume a natural parameterization of the fastest descent curve $z(s)=x(s)=x(s)+i y(s)\in\gamma_k$:
\begin{eqw}
    \frac{d}{ds}\Imag f(z(s)) = 0 \Rightarrow \Imag\left(\left(i z(s) + Z e^{iz(s)}\right)\frac{dz}{ds}\right)=0
\end{eqw}
Based on the asymptotic behavior of the constant phase curve, we can obtain two cases of the last expression:
\begin{eqw}
    \left\{
    \begin{aligned}
        &\Imag z(s) \to +\infty \Rightarrow \frac{d\Real\left(z(s)^2\right)}{ds} \approx 0 \Rightarrow \Real z(s) \approx \pm \Imag z(s) +\text{const}.\\
        &\Imag z(s) \to -\infty \Rightarrow \Imag\left(Z e^{i\Real z(s)}\frac{dz}{ds}\right) \approx 0 \Rightarrow \exists x_0: z(s) = x_0 - i\abs{\Imag z(s)}
    \end{aligned}
    \right.
\end{eqw}
Combined with the previous paragraph, this reasoning confirms that the constant phase curves in Figure \ref{saddle_points} have a general pattern that holds for any $Z$.

\subsubsection{Final contour}
\begin{figure}[!ht]
    \centering
    \includegraphics[width=\textwidth]{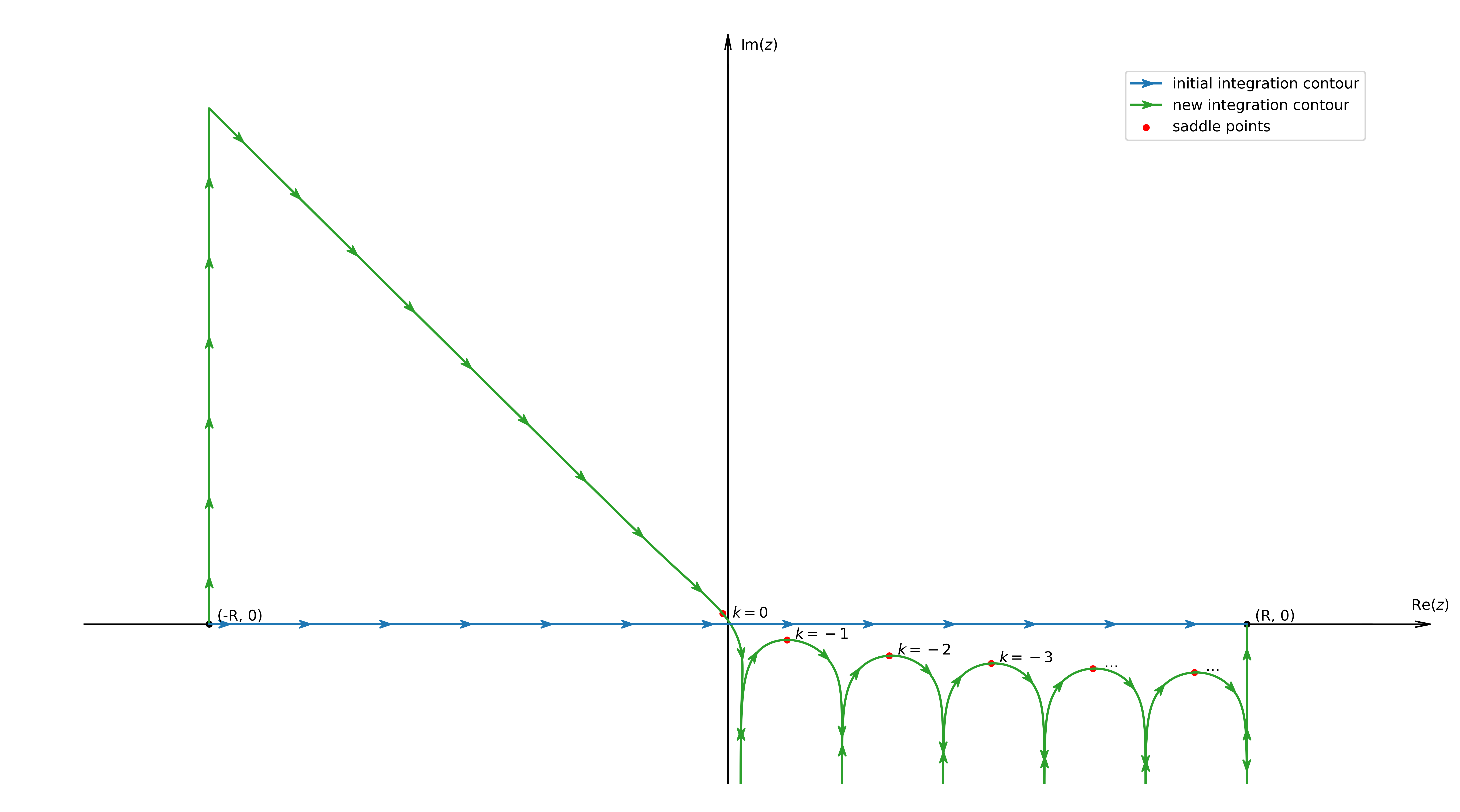}
    \caption{The blue line marks the original integration contour on the complex plane. It is simply a straight line on the real axis. The green line shows the deformed integration contour. The integrals along the green and blue contours coincide because the function $f$ has no poles. There are red dots on the green contour that indicate saddle points.}\label{figure_integration_contour}
\end{figure}

The Fig.\,\ref{figure_integration_contour} shows the proposed in \eqref{main_integral} integration contours for the case $\Gamma>0$. (For $\Gamma<0$, however, $k$ runs through nonnegative instead of nonpositive indices, resulting in a symmetric picture.) The expression \eqref{main_integral} is a principal value integral, so we can integrate from $-R$ to $R$, then take $R \to \infty$. Figure \ref{figure_integration_contour} shows how we deform the contour. In order to leave only the integrals along the constant phase curves, we prove that the integrals $I_+$ and $I_-$ along the vertical lines tend to $0$.
\begin{eqw}
    I_- &= \int\limits_{-R}^{-R+i\infty}\exp(\frac{f(z)}{2\Gamma})dz=\int\limits_{0}^{\infty}
    \exp(\frac{(-R+iy)^2}{4i\Gamma} + \frac{i Z}{2\Gamma} e^{-y-iR})idy\\
    \abs{I_-}&\leq \int\limits_{0}^{\infty} \exp(-\frac{Ry}{2\Gamma}+\frac{\abs{Z}}{2\Gamma})dy \overset{R \to \infty}{\longrightarrow}0\\
\end{eqw}
We draw the vertical line for the calculation of $I_+$ at the abscissa $R$ such that $iZ e^{iR}=-\abs{Z}<0$. It is easy to see that such $R$ are dense in the neighborhood of the point $+\infty$, which is enough to neglect $I_+$ when computing the principal value integral.
\begin{eqw}
    I_+ &= \int\limits_{R}^{R-i\infty}\exp(\frac{f(z)}{2\Gamma})dz=\int\limits_{0}^{\infty}
    \exp(\frac{(R-iy)^2}{4i\Gamma} + \frac{i Z}{2\Gamma} e^{y+iR})(-i)dy\\
    \abs{I_+}&\leq \int\limits_{0}^{\infty}
    \exp(-\frac{Ry}{2\Gamma} - \frac{\abs{Z}}{2\Gamma} e^y)dy \overset{R \to \infty}{\longrightarrow}0\\
\end{eqw}
This gives us the formula useful for further calculations:
\begin{eqw}\label{contour deformation}
    \fint\limits_{-\infty}^{\infty}\exp\left(\frac{f(z)}{2\Gamma}\right) dz = \sum\limits_{k=0}^{-\sign\Gamma \cdot \infty}
    \fint\limits_{\gamma_k} \exp\left(\frac{f(z)}{2\Gamma}\right) dz
\end{eqw}

\subsection{Residual terms by CFWW formula}
\subsubsection{Application of CFWW}
To calculate the integral along some $\gamma_k$, we will use the CFWW \cite{bleistein1975asymptotic} formula. Using this method, we need a Taylor series expansion of the function $f$ around the pass point:
\begin{eqw}\label{f_Taylor}
    f(z) = \underbrace{\frac{z_k^2}{2i} + z_k}_{f(z_k)} + \underbrace{\frac{-i-z_k}{2}}_{a_0} (z-z_k)^2 +
    \sum_{n=1}^{\infty} \underbrace{\frac{-i^n z_k}{(n+2)!}}_{a_1, \: a_2, \dots} (z-z_k)^{n+2}
\end{eqw}
To use CFWW, we mentally divide the integration contour into two parts starting at $z_k$ and going to infinity. Taking into account that the pass point has the second order, let us write down the final result:
\begin{eqw}\label{asymptotic gamma_k int raw}
    \fint\limits_{\gamma_k} \exp\left(\frac{f(z)}{2\Gamma}\right) dz
    &= \exp\left(\frac{ f(z_k)}{2\Gamma} \right)\sqrt{\frac{2\Gamma}{a_0}}\sum\limits_{n=0}^{\infty}
    \Gamma\left(n+\frac{1}{2}\right)\left(2\Gamma\right)^n c_{2n}\\
    c_{2n} &= \sum\limits_{j=0}^{2n} \frac{C_{-n-\frac{1}{2}}^j}{a_0^{n+j}}\hat{B}_{2n, j}\left(a_1, a_2, \dots, a_{2n-j+1}\right),
\end{eqw}
where $\hat{B}_{2n, j}\left(a_1, a_2, \dots, a_{2n-j+1}\right)$ are the partial (exponential) Bell polynomials of the coefficients in the Taylor series expansion of $f$ \eqref{f_Taylor}.

Bell's partial polynomials are given by the following generating function:
\begin{eqw}\label{Bell generating func}
    \exp\left(u \sum\limits_{j=1}^{\infty}x_j t^j\right) = \sum\limits_{n\geq k\geq 0}
    \hat{B}_{n,k}\left(x_1, x_2, \dots, x_{n-k+1}\right) t^n \frac{u^k}{k!}
\end{eqw}
It is easy to deduce from this definition:
\begin{eqw}\label{homogeneity of Bell polynomials}
    \hat{B}_{n, j}(\zeta\eta x_1, \zeta\eta^2 x_2, \dots, \zeta\eta^{n-j+1} x_{n-j+1}) &= \zeta^j\eta^n \hat{B}_{n, j}( x_1,  x_2, \dots,  x_{2n-j+1})
\end{eqw}
In our case, $\eta=i$ and $\zeta=-z_k$, so we only need to compute Bell polynomials of the form $\hat{B}_{2n, j}\left(\frac{1}{3!},\frac{1}{4!},\dots\right)$. These polynomials can be reduced to the well-studied associated Stirling numbers at $r=3$, which are discussed in more detail in the next subsection.

\subsubsection{Bell polynomials and associated Stirling numbers}
In this subsection we will prove the following formula:
\begin{eqw}\label{Bell and Stirling connection}
    \hat{B}_{n, k}\left(\frac{1}{r!}, \frac{1}{(r+1)!}, \dots, \frac{1}{(n-k+r)!}\right) =
    \frac{k!}{(n+(r-1)k)!}S_{r}(n+(r-1)k, k)
\end{eqw}
The right-hand side of the equation is more computationally convenient because $S_r(n, k)$ can be read by the recurrence formula\cite[~p. 222]{comtet2012advanced}
\begin{eqw}
    S_r(n+1, k) = k S_r(n, k) + C_n^{r} S_r(n-r+1, k-1)
\end{eqw}

The proof of the \eqref{Bell and Stirling connection} follows directly from a comparison of the derivative functions of Bell polynomials \eqref{Bell generating func} and the associated Stirling numbers \cite[~p. 222]{comtet2012advanced}:
\begin{eqw}\label{Stirling generating func}
    \exp\left(u\left(\frac{t^r}{r!}+\frac{t^{r+1}}{\left(r+1\right)!}+\dots\right)\right) = \sum\limits_{n=(r+1)k, k = 0}^{\infty} S_{r}(n, k) u^k \frac{t^n}{n!}
\end{eqw}

\begin{eqw}
    \sum\limits_{n\geq k\geq 0}
    \hat{B}_{n, k}&\left(\frac{1}{r!}, \frac{1}{(r+1)!}, \dots, \frac{1}{(n-k+r)!}\right) t^n \frac{u^k}{k!} =
    \exp\left(\frac{u}{t^{r-1}} \left(\frac{t^r}{r!} + \frac{t^{r+1}}{(r+1)!}+\dots \right)\right) \\
    &= \sum\limits_{n, k}^{\infty} S_{r}(n, k) \frac{u^k}{t^{(r-1)k}}\frac{t^n}{n!} =
    \sum\limits_{n, k}^{\infty} S_{r}(n+(r-1)k, k) u^k\frac{t^n}{(n+(r-1)k)!}
\end{eqw}
\subsubsection{The final expression for the sum}
First we put together the formulas \eqref{f_Taylor}, \eqref{asymptotic gamma_k int raw}, \eqref{homogeneity of Bell polynomials}, and \eqref{Bell and Stirling connection}.
\begin{eqw}
    &\Gamma\left(n+\frac{1}{2}\right) c_{2n} =
    \Gamma\left(n+\frac{1}{2}\right) \sum\limits_{j=0}^{2n} C_{-n-\frac{1}{2}}^j\frac{1}{a_0^{n+j}}
    \hat{B}_{2n, j}\left(a_1, a_2, \dots, a_{2n-j+1}\right) \\
    &\;\;\;\;\; =
    \sum\limits_{j=0}^{2n} \frac{\Gamma\left(n+\frac{1}{2}\right)\Gamma\left(-n+\frac{1}{2}\right)}{j!\Gamma\left(-n-j+\frac{1}{2}\right)}
    \frac{1}{\left(\frac{-i-z_k}{2}\right)^{n+j}}
    (-1)^n (-z_k)^j\hat B_{2n, j}\left(\frac{1}{3!}, \frac{1}{4!}, \dots, \frac{1}{(2n-j+3)!}\right)\\
    &\;\;\;\;\; =
    \sum\limits_{j=0}^{2n} \frac{\frac{\pi}{\sin\left(\pi n + \frac{\pi}{2}\right)}}{j!\frac{(-4)^{n+j}(n+j)!}{(2n+2j)!}\sqrt{\pi}}
    \frac{(-1)^n (-z_k)^j}{\left(\frac{-i-z_k}{2}\right)^{n+j}} \frac{j!}{(2n+2j)!}S_3(2n+2j, j)\\
    &\;\;\;\;\; =
    \sum\limits_{j=0}^{2n} \frac{\sqrt{\pi}}{(-4)^{n+j}(n+j)!}
    \frac{(-z_k)^j}{\left(\frac{-i-z_k}{2}\right)^{n+j}} S_3(2n+2j, j) =
    \frac{\sqrt{\pi}}{2^n(i+z_k)^n}
    \sum\limits_{j=0}^{2n} \left(-\frac{1}{2}\frac{z_k}{i+z_k}\right)^{j}\frac{S_3(2n+2j, j)}{(n+j)!}
\end{eqw}
Finally, we can substitute this into \eqref{asymptotic gamma_k int raw}, \eqref{contour deformation}, and \eqref{main_integral} to get an expression for the sum:
\begin{eqw}\label{res_sum}
    \sum\limits_{n=0}^{\infty}&\left(\frac{iZ}{2\Gamma}\right)^n  \frac{e^{i\Gamma n^2}}{n!} = e^{\frac{i\pi}{4}\sign\Gamma}
    \sum\limits_{k=0}^{-\sign\Gamma\cdot\infty} \frac{\exp\left(\frac{-i-i(z_k+i)^2}{4\Gamma}\right)}{\sqrt{-\sign \Gamma\left(i+z_k\right)}}
    \times\\
    \times&\left(1+
    \sum\limits_{n=1}^{\infty}\left(\frac{\Gamma}{i+z_k}\right)^n
    \sum\limits_{j=0}^{2n}\left(-\frac{1}{2}\frac{z_k}{i+z_k}\right)^{j}\frac{S_3(2n+2j, j)}{(n+j)!}\right),
\end{eqw}
We would like to mention the choice of the root branch. Since the original integral \eqref{main_integral} was taken within $(-\infty, \infty)$, $\gamma_k$ also has a left-to-right direction, which leads to:
\begin{eqw}
    \arg\sqrt{-\left(i+z_k\right)\sign \Gamma} \in \left(-\frac{\pi}{2}, \frac{\pi}{2}\right)
\end{eqw}

\section{Contributions from different branches of the Lambert function}\label{appB}

\subsection{Dependence on $k$}

In this section we consider the expression \eqref{res_sum} in approximate form:
\begin{eqw}
    e^{-\frac{i\pi}{4}\sign\Gamma}\sum\limits_{n=0}^{\infty}&\left(\frac{iZ}{2\Gamma}\right)^n  \frac{e^{i\Gamma n^2}}{n!} =
    \sum\limits_{k=0}^{-\sign\Gamma\cdot\infty} \frac{\exp\left(\frac{f(z_k)}{2\Gamma}\right)}{\sqrt{-\left(i+z_k\right)\sign \Gamma}}
    \left(1+O(\Gamma)\right),
\end{eqw}

Such an expression is still not convenient for computer calculations, because it contains an infinite sum. In this case, a natural question arises: which terms make the main contribution to the sum \eqref{res_sum}? Due to the smallness of $\Gamma$, the most important parameter for comparing the absolute values of the summands is $\Real f(z_k)$ as a function of $k$. When $R$ is relatively small (recall that $Z=Re^{i\Phi}$), this question is easily solved numerically, which is given by \ref{where_best_k_bar2}. However, for large $R$, we can use the decomposition for the Lambert $W$-function \cite[formula 4.20]{corless1996lambertw} and give the following expression:
\begin{eqw}\label{refzk_by_k}
    \Real(f(z_{k})) =  R\sign \Gamma + \frac{\left(\Phi + 2\pi k +  \left(R+\frac{\pi}{2} \right)\sign\Gamma\right)^2}{2R} +
     O\left(\frac{1}{R^2}\right)
\end{eqw}
It clearly shows an expression for $\bar{k}$, which is the major contributor to the sum \eqref{res_sum}:
\begin{eqw}\label{kbar}
    \bar k \approx -\left[\frac{\Phi+\left(R+\frac{\pi}{2}\right)\sign\Gamma}{2\pi}\right],
\end{eqw}
where $[...]$ means rounding to the nearest integer. The decomposition of \eqref{refzk_by_k} shows that we can often restrict ourselves to just one summand $\bar k$ of the whole sum:
\begin{eqw}\label{res_summand}
    \abs{\sum\limits_{n=0}^{\infty}\left(\frac{iZ}{2\Gamma}\right)^n  \frac{e^{i\Gamma n^2}}{n!}} =
    \frac{\exp\left(\frac{\Real f(z_{\bar k})}{2\Gamma}\right)}{\sqrt{\abs{i+z_{\bar k}}}}
    \left(1+O(\Gamma)+\exp\left(-\frac{O(\Delta \bar k)}{{R\Gamma}}\right)\right),
\end{eqw}
where $\Delta \bar k$ is defined as follows:
\begin{eqw}
    \Delta \bar k = \abs{1-2\left\{\frac{\Phi+\left(R+\frac{\pi}{2}\right)\sign\Gamma}{2\pi}\right\}},
\end{eqw}
$\Delta \bar k$ indicates how inaccurately $\bar k$ was rounded.

\subsection{Contribution decomposition by $\arg Z$}

\begin{figure}[!ht]
    \centering
    \includegraphics[width=0.75\textwidth]{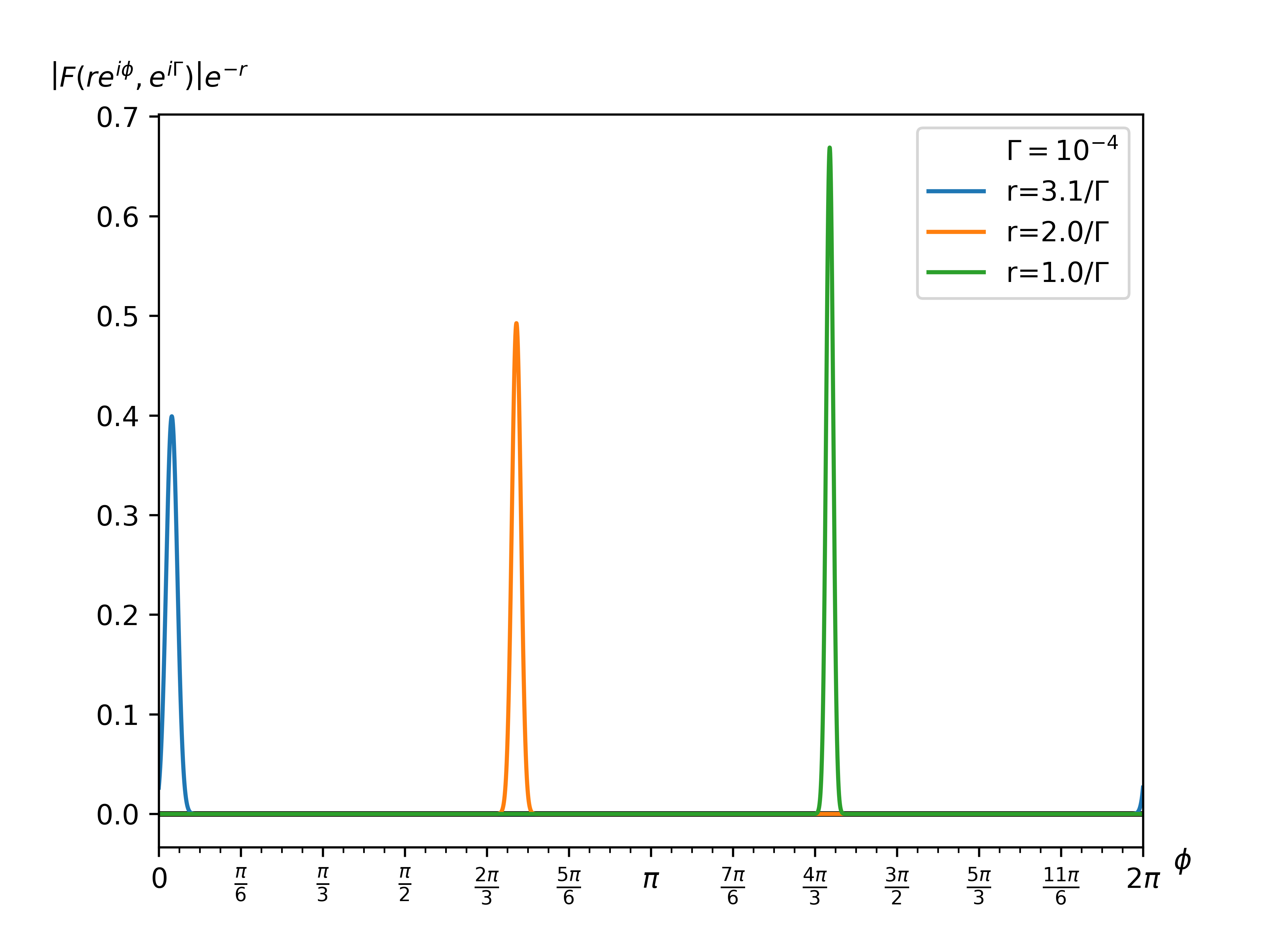}
    \caption{The representation of $F(r e^{i\phi}, \Gamma)$ as a function of the $\phi$. Due to the periodic dependence, it is sufficient to show only the dependence on the interval $[0, 2\pi)$.}\label{Fn_depending_on_phi}
\end{figure}
For large $R$, one can notice that the series \eqref{main_integral} as a function of $\Phi$ has a very narrow profile \ref{Fn_depending_on_phi}. As follows from \eqref{res_summand}, the main contribution is a multiplier of the form $\exp\left(\frac{\Real f(z_{\bar k})}{2\Gamma}\right)$. Explicitly expressing $\Real f(z_k)$ through the Lambert function,
\begin{eqw}\label{real f z k through W}
    \Real\left(f(z_k)\right) = \frac{1}{2}\Real\left(-i(z_k+i)^2\right) = -\Imag(W_k(Z))\left(1+\Real(W_k(Z))\right)
\end{eqw}
and using \cite[formula 3.2]{corless1996lambertw} we can find $\Phi_{\max}$ where the peak value is reached:
\begin{eqw}
    \frac{\partial}{\partial\Phi}\Real\left(f(z_k)\right) = -\Real(W_k(Re^{i\Phi_{\max}})) = 0
\end{eqw}
Next, from the definition of Lambert's $W$-function:
\begin{eqw}
    i\Imag W e^{i\Imag W} = R e^{i\Phi_{\max}} \Rightarrow \left|\Imag W\right| = R
\end{eqw}
The uncertainty in the sign of $\Imag W$ can be removed by considering that we are looking for a maximum, not an arbitrary extremum:
\begin{eqw}
    \sign\left(\frac{f(z_k)}{2\Gamma}\right)_{\Phi = \Phi_{max}} = 1 \Rightarrow \Imag W = - R\sign \Gamma
    &\Rightarrow \left(\frac{f(z_k)}{2\Gamma}\right)_{\Phi = \Phi_{max}} = \frac{R}{2|\Gamma|}
\end{eqw}
Using the definition of the Lambert $W$-function, we find $\Phi_{\max}$:
\begin{eqw}
    - i R \sign \Gamma  e^{- i R \sign \Gamma } = R e^{i\Phi_{\max}} &\Rightarrow \Phi_{\max} = -\left(R+\frac{\pi}{2}\right)\sign \Gamma \;\;\; \mod 2\pi
\end{eqw}
By taking higher order derivatives of \cite[formula 3.5]{corless1996lambertw}, the behavior in the neighborhood of the maximum point can be obtained:
\begin{eqw}
    \Real f(z_k) &= R\sign \Gamma + \sum\limits_{m=1}^{\infty} \frac{\left(\Phi - \Phi_{\max}\right)^{m+1}}{(m+1)!}
    \sum\limits_{k=0}^{m-1}\left\langle\left\langle{m-1}\atop {k}\right\rangle\right\rangle
    \Real\left(\frac{i^{m}\left(i R\sign \Gamma\right)^{k+1} }{\left(1-iR\sign\Gamma\right)^{2m-1}}\right)\\
    &= R\sign \Gamma\left(1 -\frac{\left(\Phi - \Phi_{\max}\right)^{2}}{2\left(1+R^2\right)}+O\left(\left(\Phi - \Phi_{\max}\right)^{3}\right)\right)
\end{eqw}
where the notation $\left\langle\left\langle{m-1}\atop {k}\right\rangle\right\rangle$ --- Euler numbers of the second kind.

\section{An alternative form of writing $F$}\label{appC}
Recall the definition of the function $F$ given in the article:
\begin{eqw}\label{Fdef}
    F(A)=\sum\limits_{m=0}^{\infty} \frac{A^m}{m!}e^{i\Gamma m^2}
\end{eqw}
In the case $\Gamma = 2\pi\frac{k}{n}$, where $\frac{k}{n}$ is an irreducible fraction with odd denominator, we can see that the multiplier $e^{i\Gamma m^2}$ runs over a finite set of values. This allows us to decompose the sum \eqref{Fdef} into $n$ sums, each of which is part of a Taylor series for the exponent. In the current section, however, we consider the approach via a functional-differential equation. In order to simplify the expressions, we assume $q=e^{i\Gamma}$.
\begin{eqw}\label{partialF}
     \frac{\partial F}{\partial A} (A) = \sum\limits_{m=1}^{\infty} \frac{A^{m-1}}{(m-1)!}q^{m^2} = \sum\limits_{m=0}^{\infty} \frac{A^{m}}{m!}q^{(m+1)^2} = qF(Aq^2)
\end{eqw}
This equation relates the derivative at point $A$ to the value at point $Aq^2$, which is on the complex plane on a circle of the same radius $\abs{A}$. Taking again the derivatives at the point $A$, we obtain:
\begin{eqw}
    \frac{\partial F}{\partial A} (A) &= qF(Aq^2) \\
    \frac{\partial^2 F}{\partial A^2} (A) &= q^{1+3}F(Aq^4) \\
    \frac{\partial^3 F}{\partial A^3} (A) &= q^{1+3+5}F(Aq^6) \\
    &\dots\\
    \frac{\partial^n F}{\partial A^n} (A) &= q^{\sum\limits_{l=1}^{n}2l-1}F(Aq^{2n}) = q^{n^2}F(Aq^{2n}) = F(A)
\end{eqw}
This is how we obtained the ordinary linear differential equation. To solve it, we need to choose $n$ different roots of $1$. Since we have chosen $q$ such that $2k$ and $n$ are coprime, the different degrees of $q^2$ completely span the entire set of $\sqrt[n]{1}$:
\begin{eqw}
    F(A) = \sum\limits_{j=0}^{n-1}C_j \exp(A q^{2j})
\end{eqw}
The equation on the constants $C_j = C_j(q)$ can be found by comparing the partial derivatives of the latter expression with \eqref{partialF}:
\begin{eqw}
    \frac{\partial F}{\partial A} (A) &= \sum\limits_{j=0}^{n-1}C_j q^{2j} \exp(A q^{2j}) = q \sum\limits_{j=0}^{n-1}C_j \exp(A q^{2j+2})\\
    C_{j+1} &= q^{-2j-1} C_j \Rightarrow C_j = q^{-j^2} C_0
\end{eqw}
The constant $C_0$ can be found from the value $F(0) = 1$, which leads to the following expression:
\begin{eqw}
     F(A) = \frac{\sum\limits_{j=0}^{n-1} q^{-j^2} \exp(A q^{2j})}{\sum\limits_{j=0}^{n-1} q^{-j^2}}
\end{eqw}

\section{Derivation of the relationship between the Wigner and Husimi functions}\label{appD}
In the main body of the article, it was mentioned that the Husimi function is related to the $F$ function as follows:
\begin{eqw}
    Q(\beta) = \frac{e^{-\abs{\alpha}^2-\abs{\beta}^2}}{\pi}\abs{F\left(\alpha\beta^*e^{-i\Gamma}\right)}^2
\end{eqw}
We use the notation $q=e^{i\Gamma}$ and $C_j(q)$ from the previous section:
\begin{equation}\label{Qnew}
	\begin{aligned}
		Q(\beta)
		&= \frac{e^{-|\alpha|^2 -|\beta|^2}}{\pi}\sum\limits_{j_1=0}^{n-1}\sum\limits_{j_2=0}^{n-1} C_{j_1}C_{j_2}^*
		\exp\left(\alpha \beta^*q^{2j_1-1}\right) \exp\left(\alpha^* \beta q^{-2j_2+1}\right)\\
		&= \frac{e^{-|\alpha|^2}}{\pi}\sum\limits_{j_1=0}^{n-1}\sum\limits_{j_2=0}^{n-1}C_{j_1}C_{j_2}^* \underbrace{
			\exp\left(-|\beta|^2 + \alpha \beta^*q^{2j_1-1} + \alpha^* \beta q^{-2j_2+1}\right)
		}_{Q^{j_1 j_2}(\beta)}
	\end{aligned}
\end{equation}
In this section, we use the following well-known relation between Husimi and Wigner functions via characteristic functions:
\begin{equation}
	\begin{aligned}
		W(\beta) = \cint \frac{d^2 z}{\pi}   e^{-i(z^*\beta + z \beta^*)} \underbrace{e^{\frac{|z|^2}{2}}
		\underbrace{\cint \frac{d^2 \tbeta}{\pi}  e^{i(z^*\tbeta + z \tbeta^*)} Q(\tbeta)}_{C_a(z)}}_{C_s(z)}
	\end{aligned}
\end{equation}

Let us sequentially compute the contribution from each $Q^{j_1 j_2}$ to $W$. The following calculations are planned as follows:
\begin{equation}
	\begin{aligned}
		C_a^\jes(z) &= \cint \frac{d^2 \beta}{\pi}  e^{i(z^*\beta + z \beta^*)} Q^\jes(\beta)\\
		W^\jes(\beta) &= \cint \frac{d^2 z}{\pi}  e^{\frac{|z|^2}{2}-i(z^*\beta + z \beta^*)} C_a^\jes(z)\\
		W(\beta) &= \frac{e^{-|\alpha|^2}}{\pi} \sum_{\jes}C_{j_1}C_{j_2}^* W^\jes(\beta)
	\end{aligned}
\end{equation}
For these calculations we need formulas, the verification of which is done by taking Gaussian integrals:
\begin{equation}
	\begin{aligned}
		\cint \frac{d^2 \beta}{\pi} \exp(-|\beta|^2 + A\beta + B\beta^*) &= \exp(A B)\\
		\cint \frac{d^2 z}{\pi} \exp(-\frac{|z|^2}{2}+i A z + i B z^*) &= 2 \exp(-2 A B)
	\end{aligned}
\end{equation}
So, let us start with the contribution of $Q^{j_1 j_2}$ to the antisymmetric characteristic function part $C_a^\jes$:
\begin{equation}
	\begin{aligned}
		C_a^\jes(z)
		&= \cint \frac{d^2 \beta}{\pi}  \exp
		\left(i(z^*\beta + z \beta^*)-|\beta|^2 + \alpha \beta^*q^{2j_1-1} + \alpha^* \beta q^{-2j_2+1}\right) \\
		&= \cint \frac{d^2 \beta}{\pi} \exp
		\left(-|\beta|^2 + (\alpha^* q^{-2j_2+1}+i z^*)\beta + (\alpha q^{2j_1-1}+i z) \beta^*\right)\\
		&= \exp\left((\alpha^* q^{-2j_2+1}+i z^*)(\alpha q^{2j_1-1}+i z)\right)\\
		&= \exp(-|z|^2 + i\alpha^* q^{-2j_2+1} z + i\alpha q^{2j_1-1}z^* + |\alpha|^2q^{-2j_2+2j_1})
	\end{aligned}
\end{equation}
then continue with its contribution to the Wigner function:
\begin{equation}
	\begin{aligned}
		W^\jes(\beta) &= e^{|\alpha|^2q^{-2j_2+2j_1}}\cint \frac{d^2 z}{\pi} \exp
		(-i(z^*\beta + z \beta^*)-\frac{|z|^2}{2} + i\alpha^* q^{-2j_2+1} z + i\alpha q^{2j_1-1}z^*)\\
		&= e^{|\alpha|^2q^{-2j_2+2j_1}} \cint \frac{d^2 z}{\pi} \exp
		(-\frac{|z|^2}{2}+i(\alpha^* q^{-2j_2+1}-\beta^* ) z + i(\alpha q^{2j_1-1}-\beta)z^*)\\
		&= 2\exp\left(|\alpha|^2q^{-2j_2+2j_1} -  2(\alpha^* q^{-2j_2+1}-\beta^* )(\alpha q^{2j_1-1}-\beta)\right)\\
		&= 2 \exp\left(-2|\beta|^2 +2\beta\alpha^* q^{-2j_2+1} + 2\beta^* \alpha q^{2j_1-1} -|\alpha|^2q^{-2j_2+2j_1}\right)\\
		&= 2  e^{-2|\beta|^2}\sum\limits_{m=0}^{\infty}\frac{(-1)^m}{m!}
		\exp(2\beta^* \alpha q^{2j_1-1})(\alpha q^{2j_1-1})^m
		\left(\exp(2\beta^* \alpha q^{2j_2-1})(\alpha q^{2j_2-1})^m\right)^*
	\end{aligned}
\end{equation}
and finally, sum all the resulting contributions:
\begin{equation}
	\begin{aligned}
		W(\beta) =& \frac{2}{\pi} e^{-|\alpha|^2-2|\beta|^2}\sum\limits_{m=0}^{\infty}\frac{(-1)^m}{m!}
		\abs{\sum_{j}C_{j}\exp(2\beta^* \alpha q^{2j-1})(\alpha q^{2j-1})^m}^2
		\\
		=& \frac{2}{\pi}  e^{-|\alpha|^2-2|\beta|^2}\sum\limits_{m=0}^{\infty}\frac{(-1)^m}{m!}
		\left|\frac{\partial^m F(2\alpha \beta^*q^{-1})}{\partial \left(2\beta^{*}\right)^m}\right|^2
	\end{aligned}
\end{equation}
The resulting expression is very well summarized through the previously introduced $F$. Let us eliminate the derivatives by reasoning similar to the derivation of the functional-differential equation \eqref{partialF}:
\begin{equation}
	\begin{aligned}
	\frac{\partial^m F(2\alpha \beta^*q^{-1})}{\partial \left(2\beta^{*}\right)^m}&=
		\left(\frac{\partial}{\partial  \left(2\beta^{*}\right)}\right)^m
		\sum\limits_{p=0}^{\infty} \frac{\alpha^p  \left(2\beta^{*}\right)^{p}}{p!}q^{p(p-1)} =
		\sum\limits_{p=0}^{\infty}  \frac{\alpha^{p+m}  \left(2\beta^{*}\right)^{p}}{p!}q^{(p+m)(p+m-1)} = \\
		&=\alpha^mq^{m(m-1)}\sum\limits_{p=0}^{\infty} \frac{(2\alpha\beta^*q^{2m-1})^p}{p!}q^{p^2} = \alpha^mq^{m(m-1)}F(2\alpha \beta^*q^{2m-1})
	\end{aligned}
\end{equation}
\begin{equation}\label{FtoWrow}
	\begin{aligned}
		W(\beta) = \frac{2}{\pi}e^{-|\alpha|^2-2|\beta|^2}\sum\limits_{m=0}^{\infty} \frac{\left(-|\alpha|^2\right)^m}{m!}\left|F(2\alpha \beta^*q^{2m-1})\right|^2
	\end{aligned}
\end{equation}

Finally, let us reduce the expression to a form containing only the Husimi and Wigner functions:
\begin{equation}\label{QtoWrow}
	\begin{aligned}
		W(\beta) = 2e^{2|\beta|^2}\sum\limits_{m=0}^{\infty} \frac{\left(-|\alpha|^2\right)^m}{m!}Q(2\beta q^{-2m})
	\end{aligned}
\end{equation}
It is important to note that this result is continuous on $q$ on the unit circle, and can be applied to any real $\Gamma$.

\section{Wigner function series}\label{appE}
\subsection{Fourier series for the function $F$}
According to the article, decomposing the Wigner function into a Fourier series is the fastest and most accurate method of calculating it.
\begin{eqw}
    \abs{F\left(\abs{A}e^{i\phi}, e^{i\Gamma}\right)}^2
    &=\sum_{m, n = 0}^{\infty} \frac{\abs{A}^{n+m}e^{i\phi(n-m)}}{m!n!} e^{i\Gamma(n^2 - m^2)} \\
    &= -\sum_{n=0}^{\infty} \frac{\abs{A}^{2n}}{n!^2} + \left(\sum\limits_{m=0}^{\infty} \sum\limits_{n=m}^{\infty}\frac{\abs{A}^{n+m}e^{i\phi(n-m)}}{m!n!} e^{i\Gamma(n^2 - m^2)} + c.c.\right)
\end{eqw}
Let us further substitute $n-m=k$. In such notations $n+m = 2m+k$.
\begin{eqw}
    &\sum\limits_{m=0}^{\infty} \sum\limits_{n=m}^{\infty}\frac{\abs{A}^{n+m}e^{i\phi(n-m)}}{m!n!}e^{i\Gamma(n^2 - m^2)} = \sum_{k=0}^{\infty}\sum_{m=0}^{\infty}\frac{\abs{A}^{2m+k}e^{i\phi k}}{m!(m+k)!}e^{i\Gamma k(2m+k)} = \\
    &=\sum_{k=0}^{\infty}e^{i\phi k}\sum_{m=0}^{\infty}\frac{\left(\abs{A}e^{i\Gamma k}\right)^{2m+k}}{m!(m+k)!} = \sum_{k=0}^{\infty}e^{ik\phi}I_{k}\left(2\abs{A}e^{i\Gamma k}\right)
\end{eqw}
Second line summation is based on modified Bessel function $I_k$. By substituting this into the formula above, we get:
\begin{eqw}\label{FsqsumI}
    \abs{F\left(\abs{A}e^{i\phi}\right)}^2 = \sum_{k=-\infty}^{\infty}e^{ik\phi}I_{k}\left(2\abs{A}e^{i\Gamma k}\right)
\end{eqw}
Using the notation $\abs{A}e^{i\Phi_0} = 2\alpha\beta^*e^{-i\Gamma}$ and substituting \eqref{FsqsumI} into \eqref{FtoWrow}, we obtain:
\begin{eqw}\label{IktoW}
    W(\beta) &= \frac{2}{\pi}e^{-|\alpha|^2-2|\beta|^2}\sum\limits_{m=0}^{\infty} \frac{\left(-|\alpha|^2\right)^m}{m!} \left|F(\abs{A} e^{i\Phi_0 + 2im\Gamma })\right|^2\\
    &= \frac{2}{\pi}e^{-|\alpha|^2-2|\beta|^2}\sum\limits_{m=0}^{\infty}\sum_{k=-\infty}^{\infty}\frac{\left(-|\alpha|^2\right)^m}{m!} e^{ik\left(\Phi_0 + 2m\Gamma \right)} I_{k}\left(2\abs{A}e^{i\Gamma k}\right)\\
    &= \frac{2}{\pi}e^{-|\alpha|^2-2|\beta|^2}
    \sum_{k=-\infty}^{\infty} e^{ik\Phi_0}I_{k}\left(4\abs{\alpha\beta^*}e^{i\Gamma k}\right)
    \exp\left(-\abs{\alpha}^2 e^{2ik\Gamma}\right)\\
    &= \frac{2}{\pi}e^{-2\left(\abs{\beta} - \abs{\alpha}\right)^2}
    \sum_{k=-\infty}^{\infty}e^{ik\Phi_0} I_{k}\left(4\abs{\alpha\beta^*}e^{i\Gamma k}\right)
    \exp\left(\abs{\alpha}^2 (1-e^{2ik\Gamma})-4\abs{\alpha\beta^*}\right)
\end{eqw}
\subsection{The required number of summands}
\subsubsection{Division into areas of $K$}
\begin{figure}[!ht]
    \centering
    \includegraphics[width=0.75\textwidth]{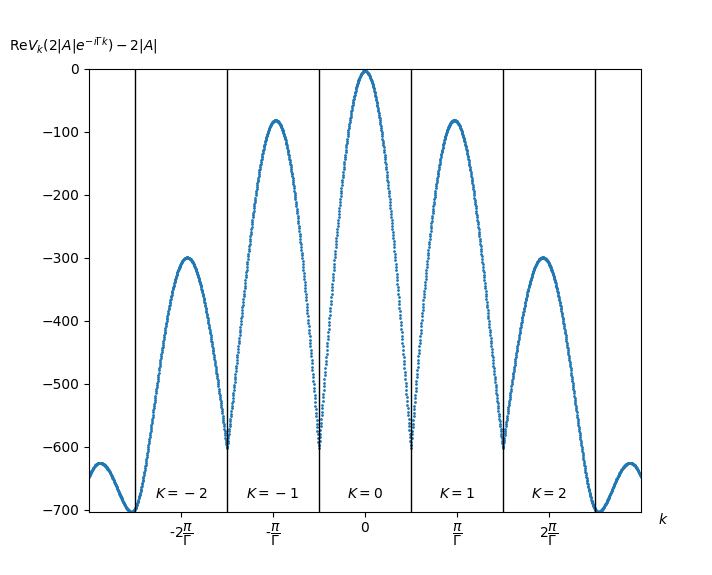}
    \caption{The typical dependence of $I_{k}\left(\abs{A}e^{i\Gamma k}\right)$ as a function of $k$. Local maxima corresponding to $e^{i\Gamma k}$ near the real axis is clearly distinguishable. The vertical lines conventionally divide the regions of $k$ belonging to different local maxima. This graph is plotted at $\abs{A}=300$, $\Gamma=0.01$.}\label{figure_ln_I_k}
\end{figure}

The equation \eqref{IktoW} is still inconvenient for direct computations on a computer. Let us write out the asymptotic expression \cite[p. 86]{bateman1953higher} for the modified Bessel function through the auxiliary $V_k$:
\begin{eqw}\label{I_bateman_asymptotic}
    V_k(z) &= \sqrt{z^2+k^2}-k\Arcsinh\left(\frac{k}{z}\sign(\Real z)\right) - \frac{1}{4}\log\left(z^2+k^2\right)\\
    I_k(z) &= \frac{1}{\sqrt{2\pi}}\exp\left(V_k(z)\right)
    \left(1+O\left(\frac{1}{\sqrt{k^2+z^2}}\right)\right)
\end{eqw}
Note that $\Real V_k(z)$ has a rather smooth dependence on $k$ and a very strong dependence on $\arg z$. $I_k(z)$ reaches local maxima when $z$ is in the neighborhood of the real axis. This is important because in the sum \eqref{IktoW} both the parameter $k$ and the phase of the argument $I_k$ change simultaneously. When $\Gamma$ is small, it is convenient to divide the regions numbered with $K$ by the number of times $4\abs{\alpha\beta^*}e^{i\Gamma k}$  occurs near the real axis. As can be seen from the graph \ref{figure_ln_I_k}, the contribution from the region $K=0$ is the largest. The contributions of the other regions will be evaluated in the next subsection.
\subsubsection{Contributions from $K\neq 0$}
To neglect contributions from regions with $K\neq 0$, the accuracy of this approximation must be determined. The function $\Real V_k(z)$ as a function of $\abs{k}$ is decreasing, and hence to evaluate the accuracy it is sufficient to compare the contributions from $K=0$ and $K=\pm 1$. Here we give the reasoning for $k = 0$ and $k = \pm\frac{\pi}{\Gamma}$ (for other $k$ the reasoning is similar, but rather cumbersome):
\begin{eqw}
     \log\abs{\frac{I_{\pm\frac{\pi}{\Gamma}}(-2\abs{A})}{I_{0}(2\abs{A})}} &=
     \Real V_{\frac{\pi}{\Gamma}}(2\abs{A}) - \Real V_0(2\abs{A}) + O\left(\abs{A}^{-1}\right)
\end{eqw}
We temporarily introduce the variable $p^{-1}=\frac{2}{\pi}\abs{A}\Gamma$ and show the difference in peak contributions:
\begin{eqw}
    \log\abs{\frac{I_{\pm\frac{\pi}{\Gamma}}(-2\abs{A})}{I_{0}(2\abs{A})}} = 2\abs{A}\underbrace{\left(\sqrt{1+p^2}-1-p\Arcsinh(p)\right)}_{<0}-\frac{1}{4}\ln\left(1+p^2\right) + O\left(\abs{A}^{-1}\right)
\end{eqw}
The function from $p$ marked by the bracket is everywhere less than zero at $p\neq 0$. Thus, the magnitude of the summands at $K\neq 0$ is exponentially suppressed.
\subsubsection{Contributions from $K=0$}
\begin{figure}[!ht]
    \centering
    \includegraphics[width=0.75\textwidth]{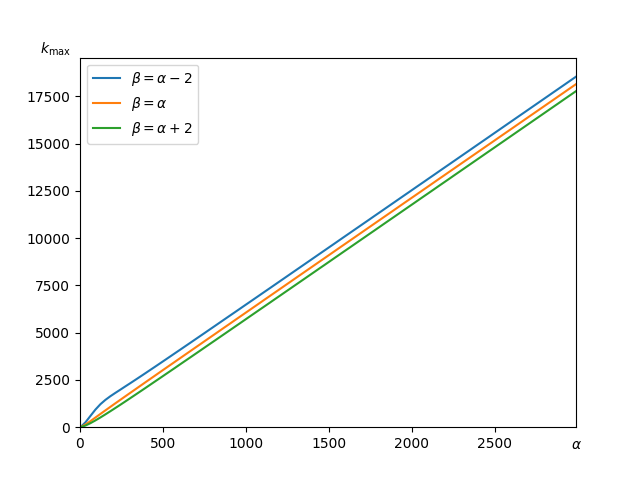}
    \caption{The picture shows the numerical dependence of $k_{\max}$ for $\beta\sim\alpha$. As already mentioned, the most interesting case is $\abs{\Gamma\alpha^2}\sim 1$. On this chart $\Gamma = \frac{2}{\alpha^2}$}
    \label{figure_k_max}
\end{figure}

In the neighborhood of $K=0$, the function $I_{k}\left(4\abs{\alpha\beta^*}e^{-i\Gamma k}\right)$ has a single maximum at $k=0$. Let us write the expression for the Wigner function:
\begin{eqw}\label{W_predfinal}
    W(\beta) \approx \frac{2e^{-2\left(\abs{\beta} - \abs{\alpha}\right)^2}}{\pi\sqrt{2\pi}}
    \sum_{k=-\frac{\pi}{2\abs{\Gamma}}}^{\frac{\pi}{2\abs{\Gamma}}}
    \exp\left(ik\Phi_0+V_k(4\abs{\alpha\beta}e^{ik\Gamma})+\abs{\alpha}^2 (1-e^{2ik\Gamma})-4\abs{\alpha\beta^*}\right)
\end{eqw}
\begin{figure}[!ht]
    \centering
    \includegraphics[width=0.75\textwidth]{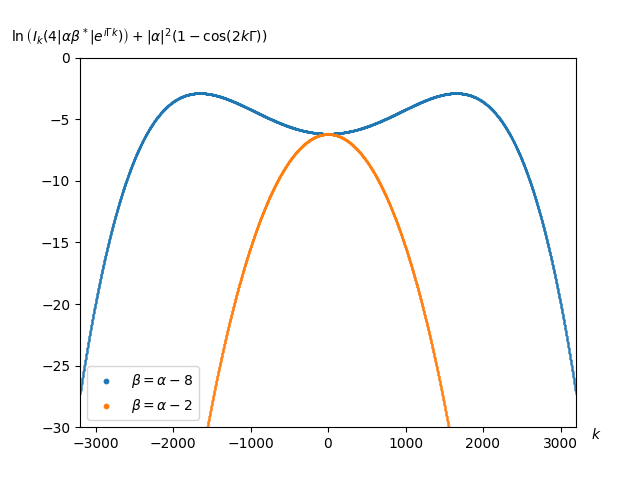}
    \caption{Picture of how logarithm of the absolute value of the summands in the \eqref{W_predfinal} expansion has two small local maxima when $\abs{\alpha}-\abs{\beta}$ is large enough. Also for comparison, more typical dependence is presented when $\beta\sim\alpha$. This graph is plotted at $\alpha=100$, $\Gamma=10^{-4}$.}
    \label{figure_ln_summand_K0}
\end{figure}
In the sum \eqref{W_predfinal} it can be seen that the modules of the terms coincide when replacing $k\rightarrow -k$. When $k>0$ the function $\Real\left(V_k(4\abs{\alpha\beta}e^{ik\Gamma})\right)$ is strictly decreasing. The addition of $\abs{\alpha}^2 (1-e^{2ik\Gamma})$ can lead to splitting the maximum into two, as shown in Figure \ref{figure_ln_summand_K0}. The condition for such splitting is the growth of the modules of the terms at $k=0$. Considering $k$ as a continuous variable, we can find $\beta_s$ for changing the sign of the second derivative at $k=0$:
\begin{eqw}
    4\abs{\alpha}\left(\abs{\alpha}-\abs{\beta_s}\right)\Gamma^2 - \frac{1}{4\abs{\alpha\beta_s}} = \frac{1}{32\abs{\alpha\beta_s}^2}
\end{eqw}
Counting $\abs{\beta_s}\sim \abs{\alpha}$, this expression can be simplified:
\begin{eqw}
    \abs{\beta_s} = \frac{\abs{\alpha}}{2}\left(1+\sqrt{1-\left(2\abs{\alpha}^2\Gamma\right)^{-2}}\right)+O\left(1\right)
\end{eqw}
Thus, at $\abs{\alpha}^2\Gamma<\frac{1}{2}$, maximum splitting does not occur, which means that the summands in the sum of \eqref{W_predfinal} will monotonically decrease at any $\beta$. With $\abs{\alpha}^2\Gamma>\frac{1}{2}$ and $\abs{\beta}<\abs{\beta_s}$, the terms will behave nonmonotonically with increasing $\abs{k}$. Since $\abs{\alpha} -\abs{\beta_s}\sim\abs{\alpha}$, we can assume that in the most meaningful area, where $\beta\sim\alpha$, figure \ref{figure_ln_summand_K0} is almost incredible in calculations. Nevertheless, the dependence of the modules of the summands is too complex to write out the upper limit $k_{\max}$ of the sum \eqref{W_final} necessary to obtain a given accuracy. In the software implementation, you can use a simple binary search to find the required $k_{\max}$. Numerical calculations show that $k_{\max} = O(\abs{\alpha})$ \ref{figure_k_max}.

%
    
\end{document}